\RequirePackage{silence}
\documentclass[aps,prl,reprint,superscriptaddress,floatfix,preprintnumbers]{revtex4-2}
\usepackage[utf8]{inputenc}
\usepackage[T1]{fontenc}
\usepackage{amsmath,amssymb}
\usepackage{graphicx}
\usepackage[dvipsnames]{xcolor}
\usepackage{booktabs}
\usepackage[colorlinks=true,allcolors=MidnightBlue]{hyperref}
\newcommand{\kT}{k_T}
\newcommand{\fNL}{f_{\rm NL}}
\newcommand{\EM}[1]{[\hyperlink{em:#1}{#1}]}
\newcommand{\anc}{\cite{Ancillary}}
\makeatletter
\def\@bibdataout@aps{\immediate\write\@bibdataout{@CONTROL{apsrev42Control\longbibliography@sw{,author="48",editor="1",pages="0",title="0",year="1"}{,author="48",editor="1",pages="0",title="",year="1"}}}\if@filesw\immediate\write\@auxout{\string\citation{apsrev42Control}}\fi}
\makeatother

\usepackage{lineno}

\linenumbers

\begin{document}
\raggedbottom

\preprint{SLAC-PUB-261110}
\title{Triple Exchange at the Cosmological Collider in Closed Form}

\author{Taushif Ahmed}
\email{taushif.ahmed@ur.de}
\affiliation{Institute for Theoretical Physics, University of Regensburg, 93040 Regensburg, Germany}
\author{Sumit Banik}
\email{banik@stanford.edu}
\affiliation{SLAC National Accelerator Laboratory, Stanford University, Stanford, CA 94309, USA}
\author{Syed Mehedi Hasan}
\email{syed-mehedi.hasan@ur.de}
\affiliation{Institute for Theoretical Physics, University of Regensburg, 93040 Regensburg, Germany}
\author{Andreas Rapakoulias}
\email{andreas.rapakoulias@ur.de}
\affiliation{Institute for Theoretical Physics, University of Regensburg, 93040 Regensburg, Germany}
\affiliation{Department of Physics, Southern Methodist University, Dallas, TX 75275, USA}

\begin{abstract}
Particles with masses near the Hubble scale leave a characteristic
pattern in the primordial bispectrum, turning inflation into a
cosmological collider. For light fields, where this signal is largest,
no closed form is known for the full shape of the quasi-single-field
triple exchange. Using Feynman-integral methods, we 
obtain the first closed-form results, exact in the momenta, as expansions
in terms of multiple polylogarithms around the massless and conformally coupled
limits. 
We show that the template used in Planck analyses overestimates the amplitude by up to a factor of 1.8 and underestimates the mass, whereas our new template recovers both, as needed for current and upcoming CMB and galaxy surveys. We also compute the collapsed trispectrum and, combining it with the bispectrum, find that at weak mixing it bounds the self-coupling more tightly.
\end{abstract}

\maketitle

\textit{Introduction}---The anisotropies of the cosmic microwave
background (CMB) and the large-scale structure of the Universe
originate from quantum fluctuations that inflation stretched to
cosmological scales. Their statistics probe physics at the
Hubble scale $H$ of inflation, up to
$10^{13}\,$GeV~\cite{Planck:2018jri}, far beyond the reach of colliders. While the power spectrum is measured to percent
precision~\cite{Planck:2018jri}, the particle content of inflation is encoded in primordial
non-Gaussianity~\cite{Maldacena:2002vr,Chen:2009zp}, most directly in
the bispectrum. In particular, a field of mass $m\sim H$ leaves a distinctive imprint in its squeezed limit, a non-analytic power law in the momentum ratio for light fields,
$m<\tfrac32H$, and an oscillation in its logarithm for heavier ones,
with an exponent or a frequency set by the
mass~\cite{Chen:2009zp,Baumann:2011nk,Assassi:2012zq,Noumi:2012vr,Arkani-Hamed:2015bza}.
This cosmological collider signal offers a unique opportunity to search for new
particles at the Hubble scale.

The search for this signal has begun. Analyses of Planck
data~\cite{Planck:2013wtn,Sohn:2024xzd,Suman:2025tpv,Kumar:2026dih,Philcox:2026tjj}
and of galaxy clustering~\cite{Cabass:2024wob,Green:2023uyz} have found no evidence so far, and the Simons
Observatory~\cite{SimonsObservatory:2018koc} and
SPHEREx~\cite{SPHEREx:2014bgr} will tighten the bounds. Since these
searches fit templates, fixed functions of the momenta whose amplitude $\fNL$
is measured, translating a bound or detection into the mass and
couplings of the new field requires the exact shape
or a calibrated template.

\begin{figure}[t]
\centering
\includegraphics[width=\columnwidth]{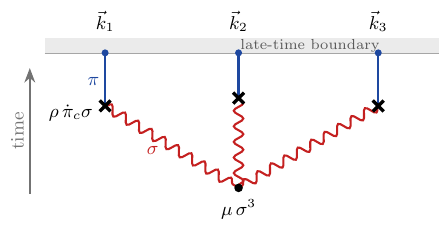}
\caption{The triple-exchange diagram, in which three massive lines (red),
coupled to inflaton legs (blue) at mixing vertices (crosses),
meet at a cubic vertex (dot).}
\label{fig:diagram}
\end{figure}

For the original model of the cosmological collider,
quasi-single-field (QSF) inflation~\cite{Chen:2009we,Chen:2009zp},
this calibration has long been missing where it matters most. In this model,
a scalar of mass $m\sim H$ mixes quadratically with the inflaton and
has a cubic self-interaction, which
at weak mixing contributes through the triple exchange of
Fig.~\ref{fig:diagram}. While the single and double exchanges are known
exactly~\cite{Arkani-Hamed:2018kmz,Qin:2023ejc,Xianyu:2023ytd,Aoki:2024uyi,Belrhali:2026act},
the latter also with high precision in the physical
region~\cite{Banik:2026cfo}, the
triple exchange is known
numerically~\cite{An:2017hlx,Werth:2023pfl,Kumar:2026dih} (see~\cite{Chen:2026iwf} for general trees), as a
one-dimensional integral~\cite{Chen:2017ryl}, as finite Mellin--Barnes (MB) integrals for heavy
fields~\cite{McCune:2026vqb}, and as hypergeometric
series~\cite{Liu:2024str,Xianyu:2025lbk,Grafe:2026avi} that diverge at physical kinematics~\cite{Xianyu:2025lbk,Grafe:2026avi}. Moreover, with the mixing resummed, it is expressed
as integrals at any mass~\cite{Huenupi:2026aqc,Pinol:2026hey,Wang:2026lff}, as series that converge
above the conformally coupled mass~\cite{Belrhali:2026uxn} and, very recently, as series in the
momentum ratios that converge at physical kinematics for any mass and
mixing~\cite{Chen:2026khu}. These representations are exact in
the mass. However, for light
fields, which decay slowly outside the horizon and give the largest signal~\cite{Chen:2009zp,Philcox:2026rpn}, the
complementary information, the momentum dependence in closed form, has not been available, and the Planck analyses~\cite{Planck:2013wtn,Sohn:2024xzd} relied on an
approximate template~\cite{Chen:2009zp} never calibrated to infer the coupling and the mass.

\begin{figure}[t]
\centering
\includegraphics[width=\columnwidth]{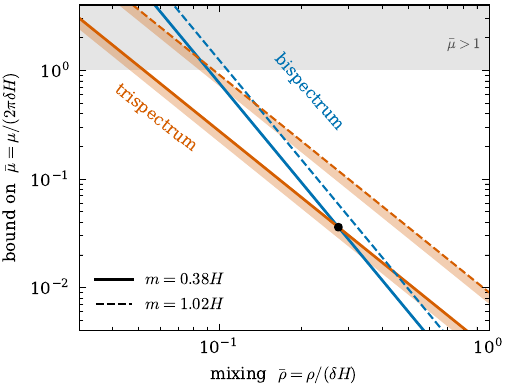}
\caption{Planck upper bounds on the self-coupling of a light field
versus its mixing, at $m=0.38H$ (solid) and $1.02H$ (dashed),
from the bispectrum (blue) and the collapsed
trispectrum~\cite{Philcox:2025wts} (orange, with a band spanning the two
component-separation pipelines). Left of each crossing (dot for $m=0.38H$), the
trispectrum gives the stronger bound.}
\label{fig:joint}
\end{figure}

In this Letter, we use methods from Feynman integral calculus to obtain,
for the first time, the QSF triple exchange of light fields as expansions in the mass that are
exact in the momenta, by two independent routes. Their coefficients are expressed in terms of multiple
polylogarithms~\cite{Goncharov:1998kja} (MPLs)
whose letters reveal the analytic structure in the momenta. The two expansions describe one function
that is meromorphic in the mass, with poles only at $m^2\le0$, and
extend to a finite duration of inflation. The QSF template,
used without calibration, biases the estimated coupling by factors up
to $1.8$ and underestimates the mass, and we provide a
simple template that measures both. Finally, the collapsed
trispectrum can separate the mixing from the self-coupling, and we combine
the Planck bispectrum and
trispectrum~\cite{Philcox:2025wts} constraints (Fig.~\ref{fig:joint}).

\textit{Setup}---In the decoupling limit of the effective field theory
of inflation~\cite{Cheung:2007st}, the canonically normalized
Goldstone boson $\pi_c$ of broken time translations couples to a
scalar $\sigma$ of mass $m$ through the QSF
interactions~\cite{Chen:2009zp}
\begin{equation}
\frac{\mathcal{L}}{a^3}\supset
\rho\,\dot\pi_c\,\sigma-\mu\,\sigma^3\,,
\label{eq:L}
\end{equation}
with $a$ the scale factor, dots denoting derivatives with respect to cosmic time, and $\rho$ and $\mu$ the
mixing and cubic couplings. In the conventions of~\cite{Werth:2024aui}, the curvature perturbation is
$\zeta=-\pi_c/(\sqrt{2\epsilon}M_{\rm pl})$, with $\epsilon$ the first
slow-roll parameter and $M_{\rm pl}$ the reduced Planck mass. At order
$\rho^3\mu$, Eq.~\eqref{eq:L} gives the diagram of
Fig.~\ref{fig:diagram}, and we compute the late-time bispectrum
$B_\pi$ in the Bunch--Davies vacuum~\cite{Bunch:1978yq}, defined by
$\langle\pi_{c,\mathbf k_1}\pi_{c,\mathbf k_2}\pi_{c,\mathbf k_3}\rangle
=(2\pi)^3\delta^{(3)}(\mathbf k_1+\mathbf k_2+\mathbf k_3)\,B_\pi$,
with $k_i=|\mathbf k_i|$. Setting $H=1$ and writing
$m^2=\tfrac94-\nu^2$, we define the shape function~\cite{Babich:2004gb}
\begin{equation}
S_\pi\equiv(k_1k_2k_3)^2B_\pi
=-\frac34\,\rho^3\mu\,k_1k_2k_3\,A_3\,,
\label{eq:S}
\end{equation}
where $A_3(k_1,k_2,k_3;\nu)$ is homogeneous of degree $-3$ in the
momenta, and the constant $-\tfrac34$ normalizes its massless limit
[see Eq.~\eqref{eq:uniform}].

In the in-in formalism~\cite{Weinberg:2005vy},
the four vertices are integrated over both branches of the
Schwinger--Keldysh contour. After Wick rotation of the conformal time,
$\eta\to\pm it$ with $t>0$ on the two branches~\cite{DiPietro:2021sjt}, two kinds of $\sigma$
lines appear~\cite{McCune:2026vqb}. An uncut line of momentum
$k$ between the times $t$ and $t'$ carries the time-ordered kernel
$K_\nu(kt_>)\,I_\nu(kt_<)$ of modified Bessel functions, whereas a cut line
factorizes into $K_\nu(kt)\,K_\nu(kt')$. Including the measure and the
external legs, a mixing vertex at time $t_l$ is integrated with
$dt_l\,t_l^{-1/2}e^{-k_lt_l}$, and the cubic vertex at time $t_0$ with
$dt_0\,t_0^{1/2}$ \EM{A}. The line of momentum $k_l$
therefore enters through one of two functions of $t_0$,
\begin{equation}
\begin{split}
L_l(t_0)&=\int_0^\infty\!\frac{dt_l}{\sqrt{t_l}}\,e^{-k_lt_l}\,
K_\nu(k_lt_>)\,I_\nu(k_lt_<)\,,\\
C_l(t_0)&=K_\nu(k_lt_0)\,,
\end{split}
\label{eq:LC}
\end{equation}
for an uncut and a cut line, with $t_>$ and $t_<$ the larger and
the smaller of $t_0$ and $t_l$. These define the four blocks of the cutting rule
of~\cite{McCune:2026vqb} (see also~\cite{Tong:2021wai}),
\begin{equation}
\begin{split}
\psi_3&=\langle L_1L_2L_3\rangle\,,\qquad
\psi_2[l]=\langle C_lL_mL_n\rangle\,,\\
\psi_1[l]&=\langle L_lC_mC_n\rangle\,,\qquad
\psi_0=\langle C_1C_2C_3\rangle\,,
\end{split}
\label{eq:blocks}
\end{equation}
where $\langle f\rangle\equiv\int_0^\infty dt_0\,t_0^{1/2}f(t_0)$,
$\{l,m,n\}$ is a permutation of $\{1,2,3\}$, and $\psi_n$ has $n$
uncut lines. Outside their convergence domain, we define these integrals by
analytic continuation in $\nu$. Cutting the line $l$ replaces
$L_l(t_0)$ by $\Phi_l\,C_l(t_0)$, where
$\Phi_l=\sqrt{\pi}\,(2k_l)^{-1/2}\Gamma(\tfrac12+\nu)\Gamma(\tfrac12-\nu)$
is the integral
$\int_0^\infty dt\,t^{-1/2}e^{-k_lt}K_\nu(k_lt)$
of the cut line over its mixing vertex. With
$s=\sin\pi\nu$, the cutting rule then gives~\EM{A}
% \begin{equation}
% \begin{split}
% A_3=&-2\psi_3+\frac{2}{\pi}(1-s)\sum_l\Phi_l\,\psi_2[l]\\
% &+\frac{4}{\pi^2}\,s(1-s)\sum_l\Phi_m\Phi_n\,\psi_1[l]\\
% &-\frac{4}{\pi^3}(1-s)^2(1+2s)\,\Phi_1\Phi_2\Phi_3\,\psi_0\,.
% \end{split}
% \label{eq:cut}
% \end{equation}
\begin{multline}
A_3=-2\psi_3+\frac{2}{\pi}(1-s)\Big[\sum_l\Phi_l\psi_2[l]
+\frac{2s}{\pi}\sum_l\Phi_m\Phi_n\psi_1[l]\\
-\frac{2}{\pi^2}(1-s)(1+2s)\,\Phi_1\Phi_2\Phi_3\,\psi_0\Big].
\label{eq:cut}
\end{multline}
With two time orderings per uncut line, $t_l<t_0$ or $t_l>t_0$, $\psi_3$, $\psi_2[l]$, $\psi_1[l]$ and $\psi_0$ give 15 integrals that we compute, while the others follow by permuting the legs.

\textit{Two routes to closed forms}---To decouple the two times in $K_\nu I_\nu$, we replace every Bessel
function by its Poisson
integral~\cite[]{NIST:DLMF},
\begin{equation}
\begin{split}
I_\nu(z)&=\frac{(z/2)^\nu}{\sqrt\pi\,\Gamma(\nu+\tfrac12)}
\int_{-1}^{1}\!d\chi\,(1-\chi^2)^{\nu-\frac12}\,e^{z\chi}\,,\\
K_\nu(z)&=\frac{\sqrt\pi\,(z/2)^\nu}{\Gamma(\nu+\tfrac12)}
\int_1^\infty\!d\lambda\,(\lambda^2-1)^{\nu-\frac12}\,e^{-z\lambda}\,,
\end{split}
\label{eq:poisson}
\end{equation}
so that in a fixed time ordering, every
exponent is linear in the times. After the
rescaling $t_l=t_0w_l$, the integral over $t_0$ is elementary and leaves
a power of $E=\sum_lk_l\xi_l$, with $\xi_l$ linear in each of the auxiliary
variables $w_l$, $\chi_l$ and $\lambda_l$. (For related Euler-type
representations, see~\cite{Katsevich:2026kgc,Grafe:2026dqn}.) To integrate over the
auxiliary variables, we separate the terms of $E$ with the MB
representation of a power of a sum~\cite{Smirnov:2012gma}. The
auxiliary integrals then factorize into Beta functions, and every
ordering becomes a multifold MB integral over a ratio of Gamma
functions and powers of the momentum ratios.

The mass enters only through $\nu$ in the Gamma
functions. With $\nu=\tfrac12-\delta$ near the conformally coupled
mass, $m^2=2$, and $\nu=\tfrac32-\delta$ in the massless limit,
$m^2=3\delta-\delta^2$, $\delta$ enters
linearly, like the regulator of dimensional
regularization. However, as $\delta\to0$, poles of the Gamma functions on
opposite sides of a contour merge and pinch it, and the integrals must
first be resolved~\cite{Czakon:2005rk,Smirnov:2009up,Tausk:1999vh,Anastasiou:2005cb}. To evaluate them, we use our package
\texttt{MBAnalytica}~\cite{MBAnalytica:2026}, which combines the
integration method of angular phase-space
integrals~\cite{Ahmed:2024pxr,Ahmed:2025yrx} with geometric methods for multifold MB
integrals~\cite{Ananthanarayan:2020fhl,Banik:2022bmk,Banik:2023rrz,Banik:2025kwa} originally developed for Feynman integrals~\cite{Smirnov:2009up,Dubovyk:2022obc}. It
resolves the $\delta$-singularities of all 15 integrals, rewrites them as Euler-type
integrals with linear denominators,
and integrates their coefficients in $\delta$ into MPLs~\cite{Vollinga:2004sn,Goncharov:1998kja} of the
momentum ratios.
% , in the notation
% $G(a_1,\ldots,a_n;z)=\int_0^zdx\,G(a_2,\ldots,a_n;x)/(x-a_1)$,
% with $G(;z)=1$ and $G(\vec 0_n;z)=\log^n (z)/n!$.

A second route starts from a conformally coupled $\sigma$,
$\nu=\tfrac12$, whose kernel is elementary,
\begin{equation}
K_{\frac12}(kt_>)\,I_{\frac12}(kt_<)
=\frac{e^{-k|t-t'|}-e^{-k(t+t')}}{2k\sqrt{tt'}}\,,
\label{eq:conf}
\end{equation}
a flat-space propagator~\cite{Arkani-Hamed:2017fdk} rescaled by
$1/\sqrt{tt'}$. Once we write the inverse powers of the times as
integrals over shifts $\omega$ of the energies,
$t^{-n-1}=\int_0^\infty d\omega\,\omega^ne^{-\omega t}/n!$, the time
integrals of every ordering are those of a flat-space tree, and every
$\psi_n$ becomes an integral, over the shifts, of the flat-space
wavefunction of the tree~\cite{Arkani-Hamed:2017fdk}, a rational
function with denominators linear in the energies. Away from $\nu=\tfrac12$, each
order in $\delta$ adds a mass insertion, a quadratic vertex
proportional to $2-m^2$ with measure $dt/t^2$, on an uncut line, or a
factor $\log(\lambda^2-1)$ under the Poisson integral of a cut line~\EM{A}, which preserves
this structure. We evaluate the resulting Euler-type integrals as in
the MB route, and the same steps apply in the massless limit, where
the kernel is also elementary.

The two routes agree where they overlap and together give all coefficients below. We also checked every
$\psi_n$ by direct numerical
integration.

\textit{Near the conformally coupled mass}---With
$\nu=\tfrac12-\delta$, the triple exchange is regular at $\delta=0$, and we obtain the Taylor coefficients $A_3^{(0)}$, $A_3^{(1)}$ and $A_3^{(2)}$ of $A_3=\sum_{n\ge0}A_3^{(n)}\delta^n$
for general kinematics as combinations of MPLs
with rational coefficients~\anc. Although $A_3$ is regular, the factors in
Eq.~\eqref{eq:cut} are not. Indeed, the factor $1-s$ vanishes at $\delta=0$, while
$\Phi_l$ and $\psi_0$ diverge. Expanding every factor in $\delta$
first, we find
\begin{equation}
A_3^{(0)}=-\frac{\pi^4}{8k_1k_2k_3}
+\sum_l\frac{\pi^2}{2k_mk_n}\,I_1[l]-2\,\psi_3^{(0)}\,,
\label{eq:A0}
\end{equation}
where the three terms come from $\psi_0$, $\psi_1[l]$ and $\psi_3$,
respectively. The function
$I_1[l]=(2/\pi)\sqrt{k_mk_n}\,\psi_1^{(0)}[l]$ has transcendental
weight two
[see Eq.~\eqref{eq:I1}], and $\psi_3^{(0)}$, a combination of 91 MPLs,
has weight four.

More generally, the coefficients of $k_1k_2k_3A_3$ at $\delta^0$,
$\delta^1$ and $\delta^2$ are pure functions~\cite{Henn:2026lfz}, with
rational coefficients and uniform transcendental weight four,
five and six. Their symbols~\cite{Goncharov:2010jf}, which keep track of the arguments, or letters, of the iterated integrals and their ordering, use the eight-letter
alphabet $\{2,k_i,\kT,\kT-2k_i\}$ of conformally coupled
fields~\cite{Arkani-Hamed:2023bsv,Arkani-Hamed:2023kig}, with
$\kT=k_1+k_2+k_3$. Hence, a coefficient can be singular only where
a letter vanishes, and each such point has a physical meaning. The point $k_i=0$ is the squeezed limit, and $\kT=0$, reached only in the unphysical region, is
dominated by early times and probes the flat-space
scattering amplitude~\cite{Arkani-Hamed:2017fdk}. The points $k_i=k_m+k_n$ are
the folded configurations, where a particle
could decay into two collinear ones, which requires a
particle in the initial state, absent in the Bunch--Davies
vacuum~\cite{Arkani-Hamed:2018kmz}. Accordingly, the letters $\kT-2k_i$
never appear as leftmost entries of the symbol, which encode the
branch points, and the functions have branch points only at $k_i=0$ and $\kT=0$.

This pattern has a simple explanation. Each massive line is $\delta$
times a twisted $d\log$ integral~\cite{Baumann:2026atn} [see
Eq.~\eqref{eq:twist}], and every vertex has the measure $dt/t$. This
predicts weight $n+4$ at order $\delta^n$, and $V+n$ for a tree with
$V$ such vertices, as anticipated for the single exchange, $V=2$, in~\cite{Baumann:2026atn}. The argument is heuristic, but we have checked it~\EM{B}, also for
two more topologies.

\textit{Light fields}---In the massless limit, with
$\nu=\tfrac32-\delta$, $A_3$ instead has a pole of fourth order at
$\delta=0$, and we compute all coefficients up to the finite part.
The pole arises because a mode of a light field decays only as
$e^{-\delta N}$ during the $N$ $e$-folds it spends outside the Hubble
radius $1/H$~\cite{Chen:2009we,Chen:2009zp}, so that each of the four
time integrals gives a factor $1/\delta$. Solving the equations of
motion in this limit~\EM{C}, we obtain the leading term, the
local shape~\cite{Komatsu:2001rj} $S_\pi=\rho^3\mu\,p_3/(4e_3\delta^4)$,
with $p_3=\sum_ik_i^3$ and $e_3=k_1k_2k_3$. With the elementary
symmetric polynomials $e_1=\kT$, $e_2=k_1k_2+k_2k_3+k_3k_1$ and
$e_3$, we write the full expansion for light fields as
\begin{equation}
\begin{split}
-3e_3^2A_3&=\mathcal{T}_\delta\sum_{n=0}^{4}q_n\,\delta^{n-4}+O(\delta)\,,\\
\mathcal{T}_\delta&=\sum_jk_j^3\prod_{i\neq j}\Big(\frac{2k_i}{\kT}\Big)^{\delta}\,,
\end{split}
\label{eq:uniform}
\end{equation}
where $q_0=1$ and $q_1=(-p_3+2e_1e_2-8e_3)/p_3$ are rational~\EM{C}, and $q_2$, $q_3$ and $q_4$ are MPLs. In the quasi-local
shape~\cite{Chen:2009we,Chen:2009zp} $\mathcal{T}_\delta$, each term $k_j^3$ comes from the power spectra of
the other two legs, proportional to $(k_ik_l)^{-3}$, and carries one
factor $(2k_i/\kT)^{\delta}$ for each of them, the decay of $\sigma$
between the horizon crossing of that mode and the cubic vertex. This
factor reproduces the exact scaling in the squeezed limit~\cite{Chen:2009zp,Baumann:2011nk},
$S_\pi\propto(k_3/k_1)^{-1+\delta}$, and resums the leading
logarithms, so that $q_2$, $q_3$ and $q_4$ remain finite as $k_3\to0$.
In the equilateral configuration,
$A_3(1,1,1;\delta)=\sum_{n=-4}^{0}c_n\,\delta^n+O(\delta)$, with
\begin{equation}
\begin{split}
c_{-4}&=-1\,,\qquad c_{-3}=-\tfrac73+2L\,,\qquad L=\log\tfrac32\,,\\
c_{-2}&=-\tfrac73+6L-2L^2+\tfrac{\pi^2}{4}-{\rm Li}_2\tfrac13\,.
\end{split}
\label{eq:cn}
\end{equation}
We obtained
$c_{-4}$ through $c_{-1}$ (see~\EM{C}) by direct integration, and $c_0$ (see~\anc) using the PSLQ
algorithm~\cite{Bailey:1999nv}.

The pole is regulated by the duration of
inflation~\cite{Chen:2009zp,Wang:2022eop}. If inflation ends $N$ $e$-folds
after the horizon crossing of the largest momentum, which we set to one,
each of the four time integrals gives $(1-e^{-\delta N})/\delta$ instead
of $1/\delta$. Matching the super-horizon evolution to the propagators at
horizon crossing~\EM{C}, we find, for $\delta\to0$ at fixed $\delta N$,
\begin{equation}
\frac{A_3(N)}{A_3(\infty)}=\big(1-e^{-\delta N}\big)^4
\Big[1+\frac{(4\Delta N-\Lambda)\,\delta}{e^{\delta N}-1}+O(\delta^2)\Big]\,,
\label{eq:finite}
\end{equation}
with elementary functions $\Delta N$ and $\Lambda$ of the momenta,
$\Delta N=0.86$ and $\Lambda=0$ in the equilateral configuration.
A massless field gives a bispectrum that grows as
$N^4$~\cite{Chen:2009zp}, and the massless and late-time limits do not
commute.

\textit{The non-Gaussian amplitude}---For
observations, we use the shape function of
$\zeta$~\cite{Babich:2004gb},
$S_\zeta\equiv(k_1k_2k_3)^2B_\zeta/(2\pi\Delta_\zeta)^4=-S_\pi/(2\pi\Delta_\zeta)$,
with $B_\zeta$ defined like $B_\pi$ and $\Delta_\zeta^2$ the
dimensionless power spectrum. In the equilateral configuration,
$S_\zeta=\tfrac{9}{10}\fNL$, so that, restoring $H$,
\begin{equation}
\fNL(m)=\frac56\,g\,A_3(1,1,1;\delta)\,,\qquad
g=\frac{\rho^3\mu}{2\pi\Delta_\zeta H^4}\,,
\label{eq:fNL}
\end{equation}
with $\delta=\tfrac32-(\tfrac94-m^2/H^2)^{1/2}$. For $m\ll H$,
$\fNL\simeq-\tfrac{5}{6}\,g\,(3H^2/m^2)^4$, and at fixed $g$ the
signal grows as $m^{-8}$, from $\fNL/g=-48$ at $m=1.02H$ to
$-1.4\cdot10^5$ at $0.38H$. Equation~\eqref{eq:fNL} applies at weak mixing,
$\rho\ll\delta H$~\cite{Chen:2009zp}, and at late times. We test
Eqs.~\eqref{eq:S} and~\eqref{eq:fNL} against
\texttt{CosmoFlow}~\cite{Werth:2023pfl,Werth:2024aui}, which computes the correlators of the theory of Eq.~\eqref{eq:L} numerically, without a mass expansion. At $m=0.38H$, for instance, our expansion for light fields agrees with \texttt{CosmoFlow} to $3\cdot10^{-6}$ in all
tested triangles. Similarly, at a finite duration, Eq.~\eqref{eq:finite} agrees with
it within $2.1\%$ at $m=0.27H$, $0.38H$ and $0.54H$ for $10\le N\le60$. For $N=60$, the
amplitude at $m=0.38H$ is $82\%$ of its late-time value.

\textit{One function of the mass}---At fixed positive momenta, $A_3$ is an even meromorphic
function of $\nu$, whose possible poles, as we show in~\EM{D},
lie only at $\nu=\pm(\tfrac32+J)$, $J=0,1,\ldots$, that is, at
$m^2/H^2=2-(J+1)(J+2)=0,-4,-10,\ldots$ (Fig.~\ref{fig:mass}). In particular, $A_3$ has no
singularity at positive $m^2$. The poles closest to the
conformally coupled mass are the massless ones, $\nu=\pm\tfrac32$,
with the Laurent coefficients
$c^{\rm L}_{-4},\dots,c^{\rm L}_{-1}$ of
Eq.~\eqref{eq:uniform} for the same triangle (the $c_n$ of
Eq.~\eqref{eq:cn} in the equilateral configuration). Subtracting
them leaves a function that is regular up to the next pole, at
$m^2=-4H^2$, so that the series in
\begin{equation}
A_3=\sum_{j=1}^4c^{\rm L}_{-j}\Big[\big(\tfrac32-\nu\big)^{-j}
+\big(\tfrac32+\nu\big)^{-j}\Big]+\sum_{n\ge0}\beta_n\,u^n\,,
\label{eq:one}
\end{equation}
with $u=\nu^2-\tfrac14=2-m^2/H^2$, converges for $|u|<6$, which
includes all light masses. With $\beta_0$, $\beta_1$ and $\beta_2$
from our closed forms near the conformally coupled mass,
Eq.~\eqref{eq:one} gives $\fNL(m)$ from the massless limit to
$\tfrac32H$, within $4\cdot10^{-4}$ of \texttt{CosmoFlow} at the
sampled masses. Once we also subtract the poles at $\nu=\tfrac52$ and
$\tfrac72$ and sum numerical coefficients of the expansion in
$\delta=\tfrac12-\nu$ through $\delta^{13}$ with
Pad\'e approximants, we continue it to heavy fields, in agreement with
\texttt{CosmoFlow} to $1.1\cdot10^{-6}$ and with the Laplace-space
solution~\cite{Belrhali:2026uxn} up to $3H$, away from $m=2.409H$,
where $\fNL$ changes sign~\EM{D}. Furthermore, expanded numerically around both
limits, the series
of~\cite{Chen:2026khu}, exact in the mass, reproduce our
coefficients.

\begin{table}[b]
\caption{Cosines $r_{XY}=F_{XY}/\sqrt{F_{XX}F_{YY}}$ of the analytic shape (E) with the templates, for the Planck 2018 temperature and
polarization Fisher matrix $F$ of \texttt{CMB-BEST}~\cite{Sohn:2023fte},
and Planck constraints on $\fNL(m)$ for E (68\% CL). The QSF template
(Q) matches E up to the rescaling $\kappa=F_{\rm QE}/F_{\rm QQ}$. The
masses are those of $\delta=\tfrac32-\nu=0.05$, $0.1$, $0.2$ and $0.4$.}
\label{tab:ovl}
\begin{ruledtabular}
\begin{tabular}{lcccccc}
$m/H$ & local & QSF & equil. & orth. & $\fNL$ & $\kappa$ \\
\colrule
0.38 & 0.995 & 0.981 & 0.35 & $-0.57$ & $-1.4\pm6.7$ & 1.82 \\
0.54 & 0.984 & 0.982 & 0.40 & $-0.60$ & $-1.7\pm8.1$ & 1.67 \\
0.75 & 0.942 & 0.985 & 0.49 & $-0.66$ & $-2.0\pm10.8$ & 1.46 \\
1.02 & 0.815 & 0.992 & 0.64 & $-0.71$ & $-0.9\pm16.2$ & 1.24 \\
\end{tabular}
\end{ruledtabular}
\end{table}

\textit{Templates and constraints}---To calibrate the
templates, we first take the squeezed limit of
Eq.~\eqref{eq:uniform}, in which the $q_n$ reduce to zeta values, and
$q_{\rm sq}(\delta)\equiv\sum_nq_n(1,1,0)\,\delta^n$ is
\begin{equation}
\begin{split}
q_{\rm sq}=\;&1+\delta+\Big(1-\frac{\pi^2}{12}\Big)\delta^2
+\Big(7-\frac{\pi^2}{12}-\frac72\zeta(3)\Big)\delta^3\\
&+\Big(27-\frac{3\pi^2}{4}-\frac72\zeta(3)-\frac{\pi^4}{6}\Big)\delta^4+O(\delta^5)\,,
\end{split}
\label{eq:Qsq}
\end{equation}
with coefficients that we identified using PSLQ. They agree with the weak-mixing limit of the squeezed
coefficient of~\cite{Huenupi:2026aqc}, which we continued
analytically to light fields. The QSF template~\cite{Chen:2009zp} used
in the Planck analyses~\cite{Planck:2013wtn,Sohn:2024xzd} and in forecasts for galaxy surveys~\cite{Sefusatti:2012ye} has the same scaling in the squeezed limit, $(k_3/k_1)^{-1+\delta}$, but, at equal
equilateral amplitude, a squeezed bispectrum lower by factors of 2.35
for $m\to0$ and 1.57 at $1.02H$ (Fig.~\ref{fig:sq}). As a result, an estimator built on this template returns $\kappa\fNL$ when the data contain our
analytic shape with amplitude $\fNL$, with $\kappa=1.82$ at
$m=0.38H$ and $1.24$ at $1.02H$ (Table~\ref{tab:ovl}). An uncalibrated bound on the
coupling $g$ is therefore too weak by
the same factor. With
cosines above $0.98$, the calibrated template has a Fisher error at most
$2\%$ larger than that of the analytic shape.

Truncating the sum in Eq.~\eqref{eq:uniform} after $q_1$ and normalizing to
$S_{\rm T}(k,k,k)=\tfrac9{10}\fNL$, we obtain a new template,
\begin{equation}
S_{\rm T}=\frac{3\fNL}{10}\,\frac{(3/2)^{2\delta}}{1+\frac73\delta}\,
\frac{\mathcal{T}_\delta}{k_1k_2k_3}\,\big(1+q_1\delta\big)\,,
\label{eq:template}
\end{equation}
an elementary function with two parameters, the amplitude $\fNL$ and,
through $\delta$, the mass. Its momentum dependence, chiefly the
slope in the squeezed limit, measures the mass, and its amplitude then gives the
coupling through Eq.~\eqref{eq:fNL}. Fitted to the analytic shape at
the masses of Table~\ref{tab:ovl}, it recovers the amplitude within 5\%
and the mass within $0.002H$. In contrast, the QSF template, fitted in the same way,
prefers lighter fields, with $m\to0$ for $0.38H$ and $0.54H$, $0.52H$ for
$0.75H$ and $0.89H$ for $1.02H$, a bias larger than the statistical
error once the signal-to-noise ratio exceeds 6 to 9.

Using the \texttt{CMB-BEST} pipeline~\cite{Sohn:2023fte}, we reproduce the
Planck template constraints~\cite{Planck:2019kim} and obtain those of the
analytic shape (Table~\ref{tab:ovl}). The
corresponding 95\% CL bound on $|g|$ ranges from $1.0\cdot10^{-4}$ at
$m=0.38H$ to $0.7$ at $1.02H$. In a simplified Fisher analysis,
the Simons Observatory~\cite{SimonsObservatory:2018koc} and a CMB-S4-like
benchmark~\cite{CMB-S4:2016ple}, combined with Planck, would reduce the error on $\fNL$ by factors of 1.6
and 2.1. In galaxy surveys, a light field also generates a
scale-dependent correction to the linear galaxy
bias~\cite{Dalal:2007cu,Sefusatti:2012ye,Gleyzes:2016tdh},
$\Delta b(k)\propto f_{\rm NL}^{\rm sq}\,k^{\delta-2}/T(k)$, with
$T(k)$ the matter transfer function. Its slope measures the mass, and,
in the normalization of local non-Gaussianity~\cite{Komatsu:2001rj},
Eq.~\eqref{eq:Qsq} gives its amplitude,
$f_{\rm NL}^{\rm sq}(m)=-\tfrac56\,g\,q_{\rm sq}(\delta)/\delta^4$, the
calibrated theoretical input for SPHEREx~\cite{SPHEREx:2014bgr} and
other surveys~\cite{Goldstein:2024bky}.

\textit{Mixing and self-coupling}---The bispectrum constrains only one
combination of the mixing and the self-coupling. In terms of
$\bar\rho=\rho/(\delta H)$ and $\bar\mu=\mu/(2\pi\delta H)$, both small
at weak coupling~\EM{C}, the coupling is
$g=\bar\rho^3\bar\mu\,\delta^4/\Delta_\zeta$, and the Planck bounds on
$g$ give $\bar\rho^{\,3}\bar\mu<7.5\cdot10^{-4}$ at $m=0.38H$ and
$1.2\cdot10^{-3}$ at $1.02H$. These bounds therefore correspond to $\bar\rho=\bar\mu\approx0.17$ at $m=0.38H$ and
$0.19$ at $1.02H$, inside the weakly coupled regime. At weak mixing, the bispectrum contains little information about
the self-coupling, since the bound on $\bar\mu$ grows as $1/\bar\rho^{\,3}$.

A second observable bounds a different combination. It is the
trispectrum from the exchange of $\sigma$ between two cubic
vertices~\cite{Assassi:2012zq,Chen:2018sce}, which depends on
$\rho^4\mu^2$. In its collapsed limit, two pairs of short modes share
one long mode of $\sigma$. In the squeezed bispectrum, the same long
mode appears only through its correlation with the inflaton, of
magnitude
$\varepsilon=\pi\rho/[2H\sin(\tfrac{\pi}{4}+\tfrac{\pi\nu}{2})]$~\cite{Assassi:2012zq}~\EM{C}.
With Eq.~\eqref{eq:Qsq}, the leading collapsed amplitude, in the
normalization of~\cite{Philcox:2025wts}, is therefore
\begin{equation}
\tau_{\rm NL}^{\rm light}=\frac{\big(\tfrac65f_{\rm NL}^{\rm sq}\big)^2}{\varepsilon^2}
=\Big[\frac{\bar\rho^{\,2}\bar\mu\,q_{\rm sq}(\delta)}{\Delta_\zeta}\,
\frac{\sin(\pi\delta/2)}{\pi\delta/2}\Big]^2\,.
\label{eq:tau}
\end{equation}
For light fields, where $\varepsilon\simeq\bar\rho$, the trispectrum is
enhanced by $1/\bar\rho^{\,2}$ relative to the squared bispectrum, and
it bounds $\bar\rho^{\,2}\bar\mu$.

The Planck analysis of the collapsed
trispectrum~\cite{Philcox:2025wts} gives
$\tau_{\rm NL}^{\rm light}<2360$ for a massless field (95\% CL), and its
result at $m=1.12H$ corresponds to $1.6\cdot10^5$. Interpolating in the
mass~\EM{C}, we estimate $\bar\rho^{\,2}\bar\mu<2.7\cdot10^{-3}$ at
$m=0.38H$ and $9.0\cdot10^{-3}$ at $1.02H$. Figure~\ref{fig:joint}
shows both bounds as upper limits on $\bar\mu$ at a given $\bar\rho$.
They fall as $1/\bar\rho^{\,3}$ and $1/\bar\rho^{\,2}$, and cross at
$\bar\rho\approx0.28$ for $m=0.38H$ and at $0.14$ for $1.02H$, where the bound is already $\bar\mu\approx0.5$. At weaker
mixing, we note that the trispectrum gives the stronger bound. For instance, at $\bar\rho=0.1$ and
$m=0.38H$, it lowers the bound from $\bar\mu<0.75$ to $0.27$. If both
signals were detected, the ratio of the trispectrum to the squared
bispectrum, $1/\varepsilon^2$, would therefore measure the mixing alone, and the
bispectrum would then give the self-coupling. In galaxy surveys, the
same enhancement also appears as a stochastic
bias~\cite{Baumann:2012bc}.

\textit{Conclusions}---With methods from Feynman integrals, we have obtained,
for the first time, the QSF triple exchange of light fields at weak mixing
in closed form in the momenta, as one meromorphic
function of the mass that agrees with the series
exact in the mass~\cite{Chen:2026khu}. The exact
shape calibrates the QSF template of the Planck analyses and
yields a simple template for the coupling and the mass.
With the collapsed trispectrum, it
can separate the mixing from the self-coupling. Our methods extend to the double exchange, to complete
trispectra and to higher orders in the mass expansion.

\textit{Note added}---During the final preparation of this Letter, the triple exchange was expressed as a convergent double series at any mass and mixing strength~\cite{Chen:2026khu}. We checked that our results numerically agree with theirs.

\textit{Acknowledgments}---The work of TA is supported by the Deutsche Forschungsgemeinschaft (DFG) through Research Unit FOR2926, \textit{Next Generation Perturbative QCD for Hadron Structure: Preparing for the Electron-Ion Collider}, project number 409651613. The work of SMH is supported by the Academic Research Sabbatical Program (ARSP) of the University of Regensburg. SB gratefully acknowledges the financial support of the Swiss National Science Foundation Project No.\ P500-2\_235325 and of the US Department of Energy under contract DE-AC02-76SF00515.

\bibliography{refs}

\newcommand{\emappendix}[1]{\setcounter{equation}{0}\renewcommand{\theequation}{#1\arabic{equation}}\renewcommand{\theHequation}{EM#1\arabic{equation}}}
\clearpage
\onecolumngrid
\section*{End Matter}
\twocolumngrid

\emappendix{A}
\hypertarget{em:A}{}\textit{A. Cutting rule and computational details}---Here we derive
the integrals of Eqs.~\eqref{eq:LC} and~\eqref{eq:blocks} and the coefficients
of Eq.~\eqref{eq:cut}. In the Euclidean time
integrals, a vertex $v$ carries $t_v^{\,r_v-1}e^{-x_vt_v}$, with $x_v$
the energy of its external legs. The power of $t_v$ collects the measure
$a^4=t_v^{-4}$, a factor $t_v^{3/2}$ per line end and, at a mixing
vertex, $t_v^{2}$ from the $\dot\pi_c$ leg, which also supplies
$e^{-k_lt_v}$. Hence $r_0=\tfrac32$ and $x_0=0$ at the cubic vertex,
and $r_l=\tfrac12$ and $x_l=k_l$ at the mixing vertices, which gives
Eqs.~\eqref{eq:LC} and~\eqref{eq:blocks}. The coefficients of Eq.~\eqref{eq:cut} come from~\cite{McCune:2026vqb}, where the correlator is a sum over the subsets
$\mathcal C$ of cut lines,
\begin{equation}
\sum_{\mathcal C}(-1)^{3-n_c}\,\frac{2}{\pi^{n_c}}\,{\rm Re}\prod_B\psi^F_B\,,
\end{equation}
with $n_c$ of the three lines cut and $\psi^F_B$ the integral of each
connected piece $B$ left by the cuts, evaluated on the uncut lines with
the Feynman kernel
\begin{equation}
G^F_\nu=K_\nu(kt_>)\,\big[I_\nu-\tfrac{i}{\pi}\,e^{i\pi\nu}K_\nu\big](kt_<)\,.
\end{equation} Its $K_\nu K_\nu$ part factorizes as on a cut line, with
the mixing-vertex integral $\Phi_l$. Collecting the terms with $n$
time-ordered lines, we obtain their weight
\begin{equation}
\sum_{j=0}^{3-n}\binom{3-n}{j}\frac{2\,(-1)^{3-j}}{\pi^{3-n}}\,
{\rm Re}\big[(s-i\cos\pi\nu)^{3-n-j}\big]\,,
\end{equation}
which evaluates to the four coefficients of Eq.~\eqref{eq:cut}.

We compute all $\psi_n$ through the finite part of $A_3$ in the massless
limit, from $\delta^{-4}$ to $\delta^{0}$, and through $\delta^{2}$ near
the conformally coupled mass. In the direct integration, an uncut line
becomes a chain of conformal kernels, Eq.~\eqref{eq:conf}, joined by mass
insertions, while a cut line is expanded in $\delta$ under its Poisson
integral, Eq.~\eqref{eq:poisson}, which gives powers of $\log(\lambda^2-1)$. We checked that the assembled $A_3$ is
symmetric under permutations of the legs to between 37 and 39 digits.
We identified the rational coefficient $q_1$ by an ansatz fitted at 28
triangles, which then holds to $10^{-15}$ at all 117 triangles we evaluated.

\begin{figure}[t]
\centering
\includegraphics[width=0.76\columnwidth]{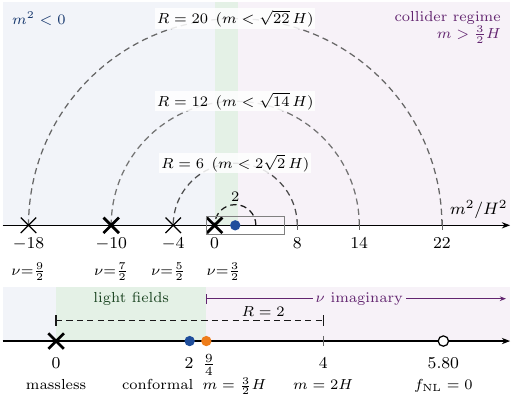}
\caption{Analytic structure of $A_3$ in the complex plane of
$m^2/H^2$. Crosses mark possible poles, of order up to four (bold) or
simple (thin). Dashed circles show the convergence of the series in
Eq.~\eqref{eq:one} as further poles are subtracted. The lower strip
magnifies the box, with the threshold $m=\tfrac32H$ (orange) and the
zero of the equilateral $\fNL$ (open circle).}
\label{fig:mass}
\end{figure}

\begin{figure}[t]
\centering
\includegraphics[width=0.8\columnwidth]{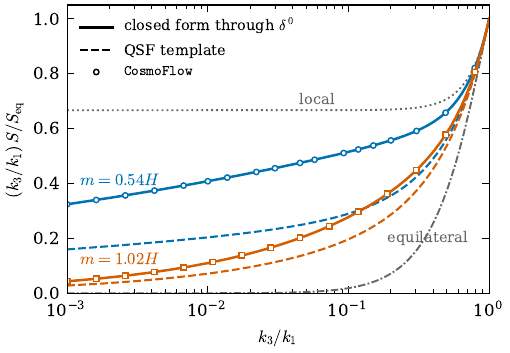}
\caption{Shape functions along isosceles triangles, $k_1=k_2$. In the
squeezed limit, the QSF template falls below our closed form for light
fields, which agrees with \texttt{CosmoFlow}.}
\label{fig:sq}
\end{figure}

\emappendix{B}
\hypertarget{em:B}{}\textit{B. Structure near the conformally coupled mass}---The simplest example of our closed forms is the weight-two function
$I_1[l]$ of Eq.~\eqref{eq:A0}. In units $k_1=1$, it reads
\begin{equation}
\begin{split}
k_l\,I_1[l]=\;&\tfrac{\pi^2}{12}+\tfrac12G(0,1;1-2k_l)
-\tfrac12G(0,1;1-\kT)\\
&-\tfrac12G\big(\tfrac1{1-\kT},\tfrac1{1-2k_l};1\big)
+\tfrac14\log^2\kT\,.
\end{split}
\label{eq:I1}
\end{equation}
The weights we find in the main text follow from the structure of the integrals. Each
Bessel function of a massive line is $\delta$ times a twisted $d\log$
integral~\cite{Baumann:2026atn},
\begin{equation}
\begin{split}
K_{\frac12-\delta}(z)=-\delta\,&\frac{\sqrt\pi\,2^{\delta-\frac12}}
{\Gamma(1-\delta)}\,z^{-\frac12-\delta}\\
&\times\int_1^\infty(\lambda^2-1)^{-\delta}\,d\log(\lambda^2-1)\,
e^{-z\lambda}\,,
\end{split}
\label{eq:twist}
\end{equation}
continued from $\delta<0$ ($I_{\frac12-\delta}$ likewise, with
$\lambda\in[-1,1]$), and the vertices of Eq.~\eqref{eq:L} have
$r_v=\tfrac12n_v$, with $n_v$ Bessel functions at the vertex. Every
vertex measure is then a twisted $dt/t$, and the time integrals give
the flat-space wavefunction~\cite{Arkani-Hamed:2017fdk} at shifted
energies, a $d\log$ form with constant coefficients. The coefficient
of $\delta^n$ is therefore pure of weight $V+n=n+4$, with leftmost
symbol entries at the physical branch
points~\cite{Arkani-Hamed:2018kmz} and letters fixed by the
$\delta$-independent singularities of the integrand. However, the argument is
heuristic, since it assumes that the nested orderings, the boundary
terms of the cut lines and the normalizations of Eq.~\eqref{eq:cut}
do not spoil this form. We therefore tested it with three checks. First, the functions, not only their
symbols, are pure at $\delta^0$, $\delta^1$ and $\delta^2$ (to 30
digits at three points). Second, every $\psi_n$ in Eq.~\eqref{eq:cut} is pure
on its own. Third, the single exchange, through fourth order, and the
three-site chain with two massive lines, at its first two orders, have
the predicted weights $n+2$ and $n+3$. The argument does not apply to light fields,
whose twist $(\lambda^2-1)^{1-\delta}$ is not a $d\log$, and indeed the coefficients in Eq.~\eqref{eq:Qsq} do not have uniform weight.

\emappendix{C}
\hypertarget{em:C}{}\textit{C. Light fields and the regime of validity}---The fourth-order
pole has a simple explanation. With $H=1$ and $N$ the number of $e$-folds since horizon crossing,
Eq.~\eqref{eq:L} gives, on super-horizon scales,
\begin{equation}
\frac{d\pi_c}{dN}=-\rho\,\sigma\,,\qquad
\frac{d\sigma}{dN}=-\frac{m^2}{3}\,\sigma-\mu\,\sigma^2\,,
\label{eq:superhorizon}
\end{equation}
where we neglect $d^2\sigma/dN^2$ and use $\tfrac{m^2}{3}=\delta+O(\delta^2)$. A light field decays only as
$\sigma_1=\sigma_*e^{-\delta N}$, with $\sigma_*$ its value at horizon
crossing, and is converted into the curvature perturbation for $1/\delta$
$e$-folds. It sources $\pi_c$ at linear order and, through its self-interaction, at
second order,
\begin{equation}
\pi_c\supset-\frac{\rho\,\sigma_*}{\delta}+\frac{\rho\mu\,\sigma_*^2}{2\delta^2}\,.
\label{eq:sourcing}
\end{equation} With
$\langle\sigma_*\sigma_*\rangle'=1/(2k^3)$, where the prime denotes
that $(2\pi)^3\delta^{(3)}(\sum_i\mathbf k_i)$ is dropped, Wick's
theorem gives the local shape
\begin{equation}
S_\pi=\frac{\rho^3\mu\,p_3}{4e_3\delta^4}\,,
\end{equation}
and hence the constant $-\tfrac34$ of Eq.~\eqref{eq:S}. In the equilateral configuration, with
$L=\log\tfrac32$ and $\ell=\log3$,
\begin{equation}
\begin{split}
c_{-1}=&-\tfrac{31}3+6L-6L^2+\tfrac43L^3+\tfrac{\pi^2}4+\tfrac{\pi^2}2L\\
&-\big(2L^2+\tfrac{\pi^2}{12}\big)\ell-(3+L)\,{\rm Li}_2\tfrac13\\
&+4\,{\rm Li}_3\tfrac23+\tfrac{\zeta(3)}{24}\,.
\end{split}
\label{eq:c0}
\end{equation}

The same picture sets the regime of validity. The late-time results
assume that inflation lasts much longer than the $1/\delta$ $e$-folds over
which $\sigma$ decays. If it ends $N$ $e$-folds after the horizon
crossing of the largest momentum, set to one, we match the super-horizon
evolution of the legs, with the conversion of $\sigma$ into $\pi_c$
accumulated until the end of inflation, to the massless integral over
the cubic vertex at horizon crossing. The matching time cancels, which
gives Eq.~\eqref{eq:finite} with
$\Lambda=\sum_jk_j^3\log(k_ik_l)/p_3$, with $\{i,l\}$ the legs other than $j$, and
$\Delta N=\tfrac56-\gamma_E-\tfrac12\log(2e_1)+(4e_1^3-9e_1e_2)/(6p_3)$.
Moreover, the matched amplitude reproduces the
first two orders of Eq.~\eqref{eq:uniform} for $\delta N\gg1$, including $q_1$, and gives
$-3e_3^2A_3=p_3[N^4+(4\Delta N-\Lambda)N^3+O(N^2)]$ for a massless field.
In the squeezed limit, where Eq.~\eqref{eq:finite} is not uniform, we
solve Eq.~\eqref{eq:superhorizon} to the end of inflation, with a common
shift of the horizon-crossing times fitted in the equilateral
configuration, and reproduce \texttt{CosmoFlow} to $0.5\%$ for
$N\ge40$. For $N=60$ and
$m\ge0.4H$, a finite duration changes the shape by at most 1.5\% for
$k_3/k_1\ge10^{-3}$, but by 3.7\% at $m=0.30H$, where $N$ must be
marginalized. Finally, the mixing is weak, and the cubic term
$\mu\sigma^2$ is a correction to the linear one, $\delta\sigma$, at
the typical amplitude $\sigma\sim H/2\pi$, for
\begin{equation}
\bar\rho=\frac{\rho}{\delta H}\ll1\,,\qquad
\bar\mu=\frac{\mu}{2\pi\delta H}\ll1\,.
\end{equation}
The linear term of Eq.~\eqref{eq:sourcing}
also gives the magnitude of the correlation of $\pi_c$ with $\sigma$ at long wavelengths,
$\varepsilon\simeq\rho/\delta$, and the late-time limit of the mixed
propagator~\cite{Chen:2017ryl} gives
$\varepsilon=\pi\rho/[2\sin(\tfrac{\pi}{4}+\tfrac{\pi\nu}{2})]$ at any
light mass. For the trispectrum, we identify the leading collapsed
term with the template of~\cite{Philcox:2025wts} and neglect subleading
terms. At $\nu=\tfrac32$ we use its 95\% limit for the local shape, with
which the template coincides. At $\nu=1$, $\tfrac12$ and $0$, we turn its
values into 95\% limits with a Gaussian likelihood and a positive flat
prior. We take the weaker pipeline and interpolate the logarithm of the
limit in $\nu$ with a cubic polynomial (a linear interpolation changes
the bounds by less than $7\%$). Equation~\eqref{eq:Qsq} is accurate to
$1.3\%$ at $\delta=0.4$.

\emappendix{D}
\hypertarget{em:D}{}\textit{D. Poles in the mass and the sign change}---The poles of
$A_3$ in the mass arise at late times. The bispectrum is one time
integral over the mixed propagators of the three
legs~\cite{Chen:2017ryl}. The Bunch--Davies initial condition of the propagators
makes its early-time part holomorphic in $\nu$, while the late-time
powers $a^p$ of the legs integrate to $a^{p+1}/(p+1)$, with
denominators that vanish only at $\nu=\pm(\tfrac32+J)$, all at $m^2\le0$. Since $A_3$ is even in $\nu$, it is a meromorphic function
of $u$. At $\nu=\tfrac32+J$ with even $J$, the triple and double cuts
contribute and the poles are of order at most four. For odd $J$, only
the double cuts survive and the poles are at most simple, with
\begin{equation}
\mathop{\rm Res}_{\nu=5/2}A_3=-\frac{\pi^2}{32}\sum_l
\frac{k_l\,[3(k_m^2+k_n^2)-k_l^2]}{(k_mk_n)^3}\,.
\end{equation}
For $m\gg H$, we integrate out the field and find
$A_3\simeq4H^6/(m^6\kT^3)$, in agreement with the equilateral limit
of~\cite{Chen:2026khu}. For positive $g$, the amplitude is thus
negative for light fields and positive for heavy ones, and it must
vanish in between, at $m=2.409H$ in the equilateral configuration
(Fig.~\ref{fig:mass}).

\emappendix{E}
\hypertarget{em:E}{}\textit{E. Origin of the template bias}---In the squeezed limit,
\begin{equation}
S_\zeta(1,1,r)\to\tfrac9{10}\fNL\,C\,r^{-1+\delta}\,,\qquad r=\frac{k_3}{k_1}\to0\,,
\end{equation}
where the slope measures the mass and $C$ fixes the ratio of the
squeezed to the equilateral signal. For $m\to0$, the exact shape is
local, with $C=\tfrac23$, whereas the QSF template~\cite{Chen:2009zp},
\begin{equation}
S\propto x^{1/2}N_\nu(8x)\,,\qquad x=\frac{k_1k_2k_3}{\kT^3}\,,
\end{equation}
with $N_\nu$ the Neumann function, has $C_{\rm QSF}\to0.284$. At
$m=1.02H$, $C=0.705$ against $C_{\rm QSF}=0.449$ (Fig.~\ref{fig:sq}).
The template thus has the correct slope but too small a squeezed
signal, the origin of both $\kappa$ and the bias toward lighter
fields. Consequently, in galaxy surveys, which measure
$f_{\rm NL}^{\rm sq}=\tfrac32C\,\fNL$, it overestimates the coupling by
$C/C_{\rm QSF}=1.6$ to $2.2$.

\end{document}